\documentstyle[aps,prl]{revtex}
\begin{document}
\input epsf.sty
\twocolumn[
\hsize\textwidth\columnwidth\hsize\csname %
@twocolumnfalse\endcsname
\draft
\widetext


\title{Neutron scattering studies of Zn-doped La$_{2-x}$Sr$_{x}$CuO$_{4}$}
 
\author{K. Hirota}
\address{Department of Physics, Tohoku University, Sendai 980-8578, Japan}

\date{\today}

{\tiny Proceeding for ISS2000}

\maketitle

\begin{abstract}
This paper reviews studies of the spatial modulation of spin in La$_{2-x}$Sr$_{x}$CuO$_{4}$ (LSCO) through
neutron-scattering measurements of the {\em elastic} magnetic peaks, with emphasis on Zn-doped LSCO.  The elastic
incommensurate magnetic peaks can be categorized in two classes with respect to the temperature dependence.  Nd-doped LSCO,
La$_{2}$CuO$_{4+y}$ and LSCO with $x \sim 1/8$ exhibit an ordinary power law behavior.  On the other hand, Zn-doped LSCO
and LSCO with $x=0.02-0.07$ show an exponential temperature dependence, and the correlation lengths stay short.  This
discrepancy can be attributed to the difference in the mechanism of pinning spin fluctuations.  The former systems have
pinning centers which is {\em coherent} to the lattice.  In the latter, however, pinning centers are randomly scattered
in the CuO$_{2}$ planes, thus {\em incoherent}.  The incoherent pinning model is discussed with referring to recent
$\mu$SR studies.

\end{abstract} 
 

\phantom{.}
]
\narrowtext
 
\section{INTRODUCTION}
\label{INTRO}


Interrelation between superconductivity and magnetism has been a central issue in the research of high $T_{c}$
superconductivity for many years\cite{Kastner_98}.  In particular, La$_{2-x}$Sr$_{x}$CuO$_{4}$ (LSCO) has been extensively
studied by neutron scattering, because LSCO has the simplest crystal structure among high-$T_{c}$ superconductors. 
The following things had been clarified by mid 1990's: (1) The mother compound La$_{2}$CuO$_{4}$ is three-dimensional
antiferromagnetic (AF) insulator\cite{Vaknin_87}. (2) The AF ordering is destroyed by a slight hole doping $(x \sim
0.02)$, and the spin glass (SG) phase appears, in which commensurate two-dimensional (2D) short-range AF fluctuations are
observed.\cite{Sternlieb_90,Keimer_92} (3) The superconductivity is realized in a certain hole concentration range $(0.05
< x < 0.27)$, and $T_{c}$ becomes maximum around $x=0.15$\cite{Keimer_92,Takagi_92,Nagao_93}. (4) Inelastic magnetic peaks
appear at four {\em incommensurate}  positions equivalent to ${\bf Q}_{\delta}=(\pi,\pi)+\delta(\pi,0)$ of the CuO$_{2}$
square unit in the superconducting range\cite{Cheong_91}: Yamada {\it et al.}\cite{Yamada_98} have shown that the
incommensurability $\delta$ follows a simple empirical relation $\delta = x$, then saturates around $\delta \sim 1/8$
beyond $x \sim 1/8$.


The inelastic incommensurate magnetic peaks in LSCO have been theoretically understood in the framework of the {\it t-J}
model as nesting of the Fermi surface\cite{Bulut_90,Tanamoto_94}.  However, this view is recently challenged by a
real-space domain picture, i.e., the stripe model\cite{Tranquada,Emery-Kivelson}.  Tranquada {\it et al.}\cite{Tranquada}
have found evidence for a static ordering of AF stripes separated by charged domain walls with an incommensurate
modulation in La$_{1.6-x}$Nd$_{0.4}$Sr$_{x}$CuO$_{4}$ (LNSCO) samples with $x=0.12$, 0.15 and 0.20. They found that the
{\em elastic} magnetic peaks appear with the incommensurability consistent with that of the dynamic spin fluctuations in
LSCO without Nd doping, and that the correlation lengths reaches $\sim 200$~\AA\ for $x=0.12$.  They have also found an
structural evidence for charge ordering.  $T_{c}$'s are reduced with respect to those in the absence of Nd doping, which
is believed to have the same origin as that in La$_{2-x}$Ba$_{x}$CuO$_{4}$ near $x=1/8$\cite{Moodenbaugh_88,Axe_94}
because both systems show a transition to low temperature tetragonal (LTT) phase below $\sim 70$~K in contrast with LSCO
which retains the low temperature orthorhombic (LTO) structure to the lowest temperature measured.  Tranquada {\it et
al.}\cite{Tranquada} speculated that the LTT phase is essential for the static stripe order, which competes with the
superconductivity.


The coincidence between the elastic and inelastic magnetic peak positions has led to reconsideration of static and dynamic
spatial structures of spin in high $T_{c}$ superconductors, particularly from the point of view of the stripe
model.  In this paper, review is given of the results of a wide variety of neutron-scattering studies on the elastic
incommensurate magnetic peaks in Sec.~\ref{OVERVIEW}.  Several important features concerning spatial structures of spin
are summarized in Sec.~\ref{SPACE}.  We then categorize the elastic magnetic peaks into two classes with respect to the
pinning mechanism of AF fluctuations, namely, coherent and incoherent pinning. 
 
\section{Overview of the elastic magnetic peaks in LSCO}
\label{OVERVIEW}

It was a fundamental issue whether the static magnetic order found in LNSCO is its characteristic feature or somewhat
universal to high-$T_{c}$ superconductors.  It is known that $T_{c}$ is suppressed by Zn substitution for Cu,  and also in
LSCO with $x \sim 1/8$.  Goto {\it et al.}\cite{Goto_94} have studied LSCO with $x=0.115$ by NMR and concluded that
magnetic order appears even in the absence of the LTT structure.  Kumagai {\it et al.}\cite{Kumagai_94} have clarified by
$\mu$SR measurements that there exists a static magnetic field in LSCO, which onset temperature reaches maximum at
$x \sim 0.12$.  Since the magnetic ordering seems more stable with lower $T_{c}$ samples in LNSCO, it was natural to start
looking for elastic magnetic signals in Zn-doped LSCO and LSCO with $x \sim 1/8$.

\begin{figure}
\centerline{\epsfxsize=3.25in\epsfbox{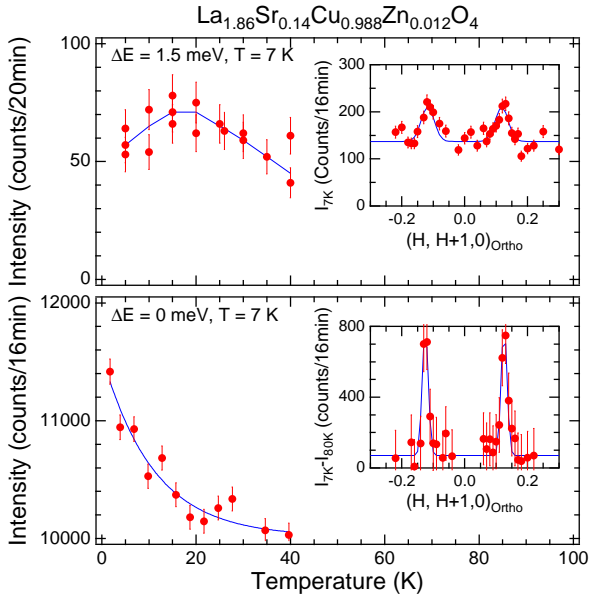}}
\vspace*{0.05in}
\caption{Temperature dependence of the incommensurate magnetic peaks at 0 and 1.5~meV transfer for
La$_{1.86}$Sr$_{0.14}$Cu$_{0.988}$Zn$_{0.012}$O$_{4}$ ($T_{c}=19$~K).  Each inset shows a peak profile along orthorhombic
$(q\ 1+q\ 0)$ at 7~K.  From Hirota {\it et al.}\protect\cite{Hirota_98}.}
\label{Fig:Zn}
\end{figure}

Hirota {\it et al.}\cite{Hirota_98} have observed quasielastic magnetic peaks, with the correlation length of $\sim
80$~\AA\, in La$_{2-x}$Sr$_{x}$Cu$_{1-y}$Zn$_{y}$O$_{4}$ (LSCZO) with $x=0.14$ and $y=0.012$ ($T_{c}=19$~K).  As shown in
Fig.~\ref{Fig:Zn}, the quasielastic component starts increasing below $T_{m} \sim 20$~K, while the low-energy inelastic
response appears at the same $\delta$ with a broad intensity maximum around $T_{m}$.  These findings indicate that small
Zn doping induces quasistatic AF ordering at low temperature by shifting the spectral weight from inelastic region. 
These results mirrored the behaviors seen in samples in the SG region\cite{Sternlieb_90,Keimer_92}


Suzuki {\it et al.}\cite{Suzuki_98} have obtained evidence for incommensurate magnetic order at low temperature in LSCO
with $x=0.12$ by using a double-axis neutron diffractometer.  Detailed elastic neutron-scattering measurements were carried
out by Kimura {\it et al.}\cite{Kimura_99,Kimura_00}.  They have found {\em elastic} incommensurate magnetic peaks
indicating a spin-density wave (SDW) order with correlation length exceeding 200~\AA.  The magnetic peaks start appearing
below $T_{m} \sim 30$~K, which coincides with onset $T_{c}=31$~K.  Subsequent studies by Matsushita {\it et
al.}\cite{Matsushita_99} on LSCO with $x=0.10$ and 0.13 have revealed that $T_{m}$ rapidly decreases and differs from
$T_{c}$, and that the correlation lengths also drastically diminishes.  For the $x=0.15$ and 0.18 samples, Yamada {\it et
al.}\cite{Yamada_98} had already confirmed that there exists no elastic magnetic scattering and that only inelastic
incommensurate scattering remains with an energy gap of $\sim 4$~meV at low temperature.  

\begin{figure} 
\centerline{\epsfxsize=3.25in\epsfbox{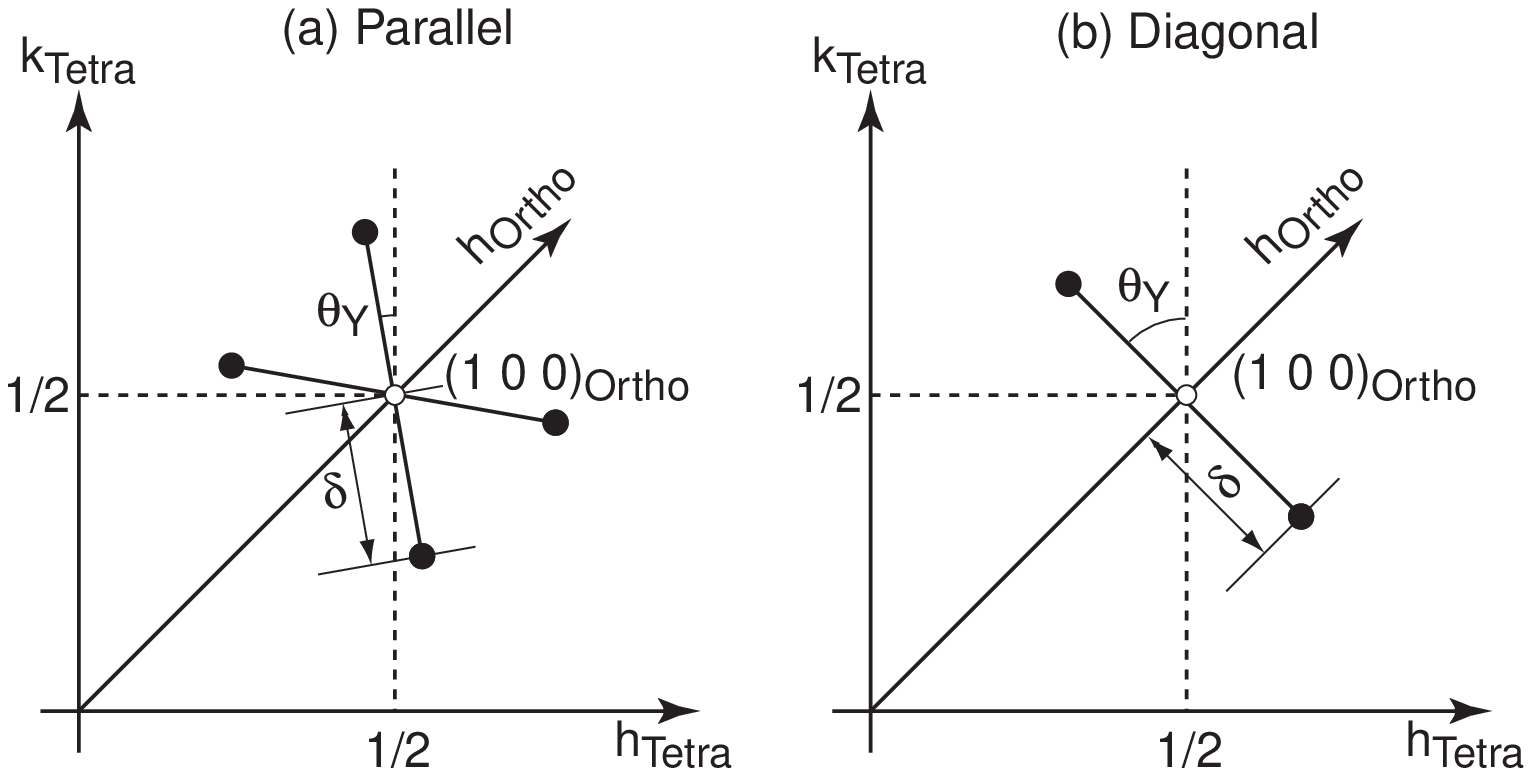}}
\vspace*{0.05in}
\caption{Schematic diagram of the incommensurate positions of magnetic peaks (a)
parallel and (b) diagonal to the Cu-O-Cu direction.  Note that the angle $\theta_{Y}$ is defined as a
rotation from the tetragonal $[0\ 1\ 0]$ direction.}
\label{Fig:Diagram}
\end{figure}

These discoveries are important not only because the incommensurate SDW order is confirmed in the absence of LTT,
but also because they have urged efforts to understand the spatial structure of spin in further detail than that
accomplished in inelastic neutron-scattering studies.   Lee {\it et al.}\cite{Lee_99} have found a long-range
SDW order in the superconducting state of a predominantly stage-4 La$_{2}$CuO$_{4.12}$ ($T_{c}=42$~K). 
They have found that the elastic magnetic peaks appear at the same temperature within the errors as the superconductivity,
suggesting that the two phenomena are strongly correlated.  Even more importantly, as shown in Fig.~\ref{Fig:Diagram}(a),
their detailed measurements have revealed that the elastic magnetic peaks are not located at
4-fold symmetrical position around $(\pi,\pi)$, i.e., corners of a square, but form a rectangular which has 2-fold
symmetry around the orthorhombic $[1\ 0\ 0]$ axis.  This indicates that the direction of the spin modulation is
rotated from perfect alignment along the Cu-O-Cu direction by a finite angle $\theta_{Y}$, which is 3.3$^{\circ}$ for
La$_{2}$CuO$_{4.12}$.  The shift of SDW peaks, or the $Y$ shift, implies that the magnetic correlations have
one-dimensional anisotropy on the 2D CuO$_{2}$ plane.  The $Y$ shift was also confirmed in LSCO
$x=0.12$\cite{Kimura_00}.


The magnetic scattering in the SG phase of LSCO ($0.02 \le x \le 0.05)$ was also revisited.  One of the most important
findings was done by Wakimoto {\it et al.}\cite{Wakimoto_99}, who have discovered new satellite magnetic peaks in LSCO
$x=0.05$ at positions rotated by $\sim 45^{\circ}$ in reciprocal space about $(\pi,\pi)$ from those found in the
superconducting phase ($x \ge 0.06$).  Due to twinning in LSCO single crystals, the orthorhombic $(1\ 0\ 0)$ and $(0\ 1\
0)$ positions nearly coincide with each other at $(\pi,\pi)$.  Wakimoto {\it et al.}\ have employed a narrower instrumental
$Q$ resolution and succeeded in ascribing a certain magnetic peak to a certain twin; in a single twin, there exist only
two satellites with the modulation vector along the orthorhombic $b^{*}$ axis as shown in
Fig.~\ref{Fig:Diagram}(b)\cite{Wakimoto_00}.  These works indicate that the SDW modulation is diagonal to the Cu-O-Cu axis
for the $x=0.05$ insulating sample, and suddenly turns to be parallel to the direction upon crossing the
insulator-to-superconductor phase boundary at $x_{c} \sim 0.055$.   Subsequent studies  by Fujita {\it et
al.}\cite{Fujita_00} have revealed that the well-defined diagonal peaks in the insulating region suddenly broadens upon
entering the superconducting phase and finite intensities appear at the parallel positions, which indicate an intimate
relation between the parallel SDW modulation and the superconductivity.   Lower hole concentration region has been studied
by Matsuda {\it et al.}\cite{Matsuda_00,Matsuda_00_2}.  They have found that the diagonal SDW modulation occurs universally
across the insulating SG phase down to $x=0.02$.

\section{Spatial structures of spin in LSCO}
\label{SPACE}

Table~\ref{Table:Summary} summarizes characteristic properties of elastic magnetic scattering observed in LSCO and related
compounds to date.   Several important features of the elastic magnetic peaks are listed as follows:

\begin{itemize}
\item The elastic peaks in the superconducting phase appears at nearly the same incommensurate position for the inelastic
peaks.  Shift of spectral weight from inelastic to elastic component with lowering temperature was observed in
LSCZO\cite{Hirota_98}.

\item The elastic peaks in the superconducting phase exhibit a characteristic rotation, called {\it Y} shift\cite{Lee_99}
around $(\pi,\pi)$.

\item Transition from Parallel to Diagonal SDW coincides with the appearance of superconductivity at $x_{c} \sim
0.055$\cite{Wakimoto_00,Fujita_00}.

\item Incommensurability $\delta$ follows an empirical relation $\delta = x$ for $x \le 1/8$, even below $x_{c}$. It
deviates from the line below $x=0.024$\cite{Matsuda_00_2}.

\item The correlation lengths $\xi$ drastically shorten upon doping.  $\xi$ shows a large anisotropy in low
$x$\cite{Matsuda_00}, but becomes isotropic with increasing $x$.  $\xi$ appears shortest near
$x_{c}$\cite{Fujita_00}.

\item No significant anomaly is observed at $x_{c}$ in the $x$ dependence of the magnetic moment
$\mu$\cite{Wakimoto_00_2} as observed for the orthorhombicity\cite{Fujita_00}.

\item $\xi$ is exceptionally long in LSCO $x \sim 0.12$\cite{Kimura_00}, LNSCO\cite{Tranquada}, and
La$_{2}$CuO$_{4+y}$\cite{Lee_99}.

\item $T_{m}$ seems enhanced where $T_{c}$ is suppressed as seen in LSCO and LNSCO, which seems however not the case for
La$_{2}$CuO$_{4+y}$\cite{Lee_99}.
\end{itemize}

The spatial spin structure on the CuO$_{2}$ plain is closely related to the superconductivity. 
However, the SDW order seems not completely universal in all LSCO and related compounds.

\section{Coherence of Pinning}
\label{PIN}

\begin{figure}
\centerline{\epsfxsize=3.25in\epsfbox{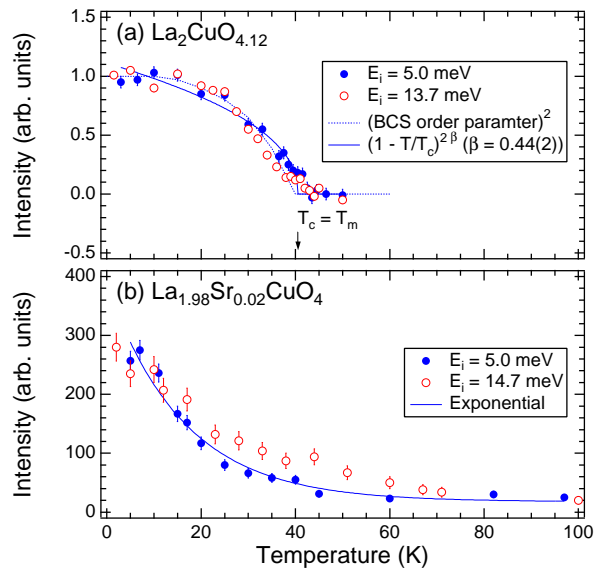}}
\vspace*{0.05in}
\caption{(a) Temperature dependence of the peak intensity of the incommensurate elastic
scattering of La$_{2}$CuO$_{4.12}$.  Filled and open circles correspond to
the incident energy $E_{i}$ of 5.0 and 13.7~meV, yielding energy resolutions of 0.15 and
0.9~meV FWHM. From Lee {\it et al.}\protect\cite{Lee_99}. (b) Temperature dependence for LSCO with $x=0.02$.
Filled and open circles correspond to $E_{i}$ of 5.0 and 14.7~meV, yielding 0.25 and 0.9~meV FWHM resolutions. From
Matsuda {\it et al.}\protect\cite{Matsuda_00}.}
\label{Fig:Tdep}
\end{figure}

Let us examine the temperature dependence of the elastic magnetic peaks.  As shown in Fig.~\ref{Fig:Tdep}, the order
parameter exhibits an ordinary power-law behavior for La$_{2}$CuO$_{4.12}$\cite{Lee_99}, but exponentially increases with
decreasing temperature for LSCO with $x=0.02$\cite{Matsuda_00}.  As a matter of fact, all the elastic magnetic scattering
peaks studied can be categorized into these two classes, power-law and exponential.  LNSCO with $x=0.12$ and
$x=0.15$\cite{Tranquada}, LSCO with $x=0.12$\cite{Kimura_99} and La$_{2}$CuO$_{4+y}$ show a power-law ordering, while LSCO
with $x=0.2-0.7$\cite{Wakimoto_99} and Zn-doped LSCO\cite{Hirota_98,Kimura_99} display a more like exponential curve.  The
magnetic moment in LNSCO with $x=0.20$ is small and overwhelmed by Nd-ordering, thus the shape of the order parameter is
not very clear.


If the elastic magnetic component is derived from the inelastic components as mentioned, there should be some specific
mechanisms condensing AF spin fluctuations.  In the LTT structure, the tilt direction of the CuO$_{6}$ octahedra is along
the Cu-O-Cu direction, i.e., tetragonal $[1\ 0\ 0]$, which is $\sim 45^{\circ}$ rotated from $[1\ 1\ 0]$ as in the LTO
phase.  Tranquada {\it et al.}\cite{Tranquada} speculated that a coupling between the tilt modulation and the charge-stripe
correlations is possible only when the tilts have a $[1\ 0\ 0]$ orientation, which induces the SDW order at lower
temperature.  As for the charge-stripe order, their speculation concerning the necessity of LTT seems valid because no
signature of the charge ordering has been reported for LTO systems.  However, hole concentration near
$x \sim 1/8$ and excess oxygens which show staging order in La$_{2}$CuO$_{4+y}$ can also induce a long-range SDW order.  A
feature common to LTT, $x \sim 1/8$ and oxygen staging order is {\em coherence} to the lattice, though
the coherence length may be shorter for $x \sim 1/8$ than for the others.  They are pinning centers which periodicity is
commensurate with the lattice, and may change the electronic and structural state of the CuO$_{2}$ plane in a fairly long
range.  In these cases, the magnetic correlations can be developed to a true long-range order.

On the other hand, doped Zn ions and holes in low density are randomly distributed in the CuO$_{2}$ plane, thus {\em
incoherent} to the lattice.  These pinning centers slow AF fluctuations locally and form local SDW domains. 
Considering the correlation length estimated for LSCZO with $x=0.14$ and $y=0.012$, $\xi \sim 80$~\AA\ , one might
realize that $\xi$ is too long for each SDW domain around a Zn ion to be separated from the other domains because the
average distance between nearest Zn ions is $\sqrt{a^{2}/y} \sim 35$~\AA\ where $a$ is the Cu-O-Cu distance and $y$ is the
doping rate of Zn.  If magnetic order induced around Zn is static and has more or less similar spatial spin modulation,
most of the domains are connected or {\em percolated} and result in a long-range order.

Nachumi {\it et al.}\cite{Nachumi_96} have performed transverse-field $\mu$SR measurements of Zn-doped LSCO.  They have
estimated the superconducting carrier density/effective mass $n_{s}/m^{*}$ ratio at low temperature, and found that
$n_{s}/m^{*}$ decreases with increasing Zn concentration.  They have then proposed ``swiss cheese'' model where charge
carriers within an area $\pi\xi_{\mu{\rm SR}}^{2}$ around each Zn are excluded from superconductivity.  For  LSCZO with
$x=0.15, y=0.01$ and $x=0.20, y=0.01$, they have obtained $\xi_{\mu{\rm SR}}=18.3$~\AA.  For the $x=0.15, y=0.01$ sample,
weak static magnetic order was confirmed below 5~K.  

Their ``swiss cheese'' model seems relevant to our local and incoherent SDW domain picture, though the characteristic
lengths are quite different.  One of the major difference between $\mu$SR and neutron scattering is their time scale. 
``Elastic'' means fluctuations slower than 10$^{11}$Hz (0.4~meV) for many thermal neutron-scattering studies, while
$\mu$SR covers 10$^{7}$ to 10$^{10}$~Hz for observable fluctuating phenomena.  This difference in time scale explain the
discrepancy in the magnetic ordering temperatures of LSCZO, 30~K for neutron scattering and 5~K for $\mu$SR.  This picture
is consistent with the view that the magnetic signals is quasielastic rather than truly elastic so that the temperature
dependence of the intensity depends on the energy window as shown in LSCO with $x=0.02$ (Fig.~\ref{Fig:Tdep}(b)).  This
{\em window} dependence is not observed in La$_{2}$CuO$_{4+y}$, though very slow critical divergence with temperature
may cause such dependence even in the coherent pinning case.  Kimura {\it et al.}\cite{Kimura_99} have pointed out that
doping Zn to LSCO with $x \sim 0.12$ considerably shortens the magnetic correlation length, 200~\AA\ to 77~\AA.  This
implies that the incoherent pinning can be destructive to the coherent pinning.  On the other hand, the SDW order is most
stable at $\delta \sim 1/8$ for LNSCO (LTT), which indicates that two different kinds of coherent pinning mechanisms are
cooperative.

\begin{figure} 
\centerline{\epsfxsize=3.25in\epsfbox{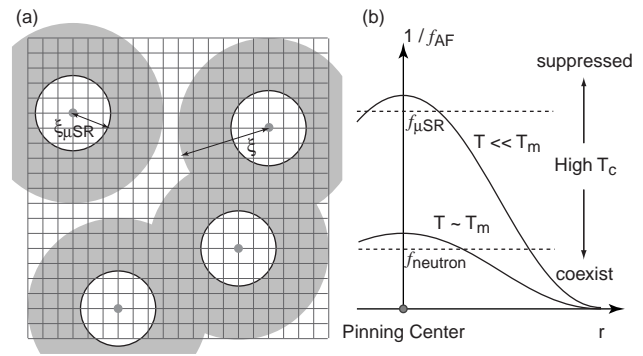}}
\vspace*{0.05in}
\caption{(a)  Schematic representation of quasistatic SDW domains around Zn ions for LSCZO
with $y \sim 0.01$.  $\xi \sim 23$~\AA\ and $\xi_{\mu{\rm SR}} \sim 10$~\AA\ are assumed for clarification.  (b) Spatial
dependence of the inverse of the characteristic AF fluctuations, $1/f_{AF} (r)$, around one of pinning centers, which are
randomly distributed, at $T
\le T_{m}$ and $T \ll T_{m}$.}
\label{Fig:Model}
\end{figure}

Let us assume that the characteristic frequency of AF fluctuations has spatial dependence around Zn, as schematically drawn
in Fig.~\ref{Fig:Model}.  Namely AF fluctuations are slowed down more drastically as closer to Zn and as temperature is
lowered.  Slightly below $T_{m}$ determined from neutron scattering, the AF fluctuations near Zn can be detected by
{\em elastic} neutron scattering, but not by $\mu$SR.  Indeed, the intrinsic energy-width, i.e, the characteristic
frequency, of the magnetic elastic peaks was estimated for LSCZO with $x=0.14$ and $y=0.012$, which is $\Gamma = 0.22$~meV
$\sim 0.05$~THz.\cite{Hirota_98}  At lower temperature, $\mu$SR starts detecting magnetic ordering from a region smaller
than that for neutron scattering.  In this model, the edge of the SDW domain is still substantially fluctuating, thus
overlap of the domains may not necessarily result in percolation of magnetic coherence.  Although it is not clear whether
$\xi_{\mu{\rm SR}}$ calculated for {\em non-superconducting} region is the same for magnetic-ordering region in $\mu$SR
time scale, these two entities are most likely correlated.   Nevertheless, the model implies that there is a characteristic
frequency of AF fluctuations for superconductivity, $f_{c}$: Superconductivity is suppressed in the region where
AF fluctuations are slower than $f_{c}$.   It is, however, not clear whether AF fluctuations with particular spatial
modulation and frequency faster than $f_{c}$ are necessary to or just compatible to high-$T_{c}$ superconductivity.


Systematic studies by combining probes with different time scales are clearly required to elucidate the nature of
the {\em quasistatic} SDW ordering as well as the role of Zn doping in magnetism and superconductivity.

\acknowledgments 

The author gratefully acknowledges R.J. Birgeneau, Y. Endoh, M. Fujita, H. Hiraka, M.A. Kastner, H. Kimura, Y.S. Lee, G.
Shirane, S. Wakimoto, and K. Yamada for international collaborative works on neutron scattering of the elastic
magnetic peaks in LSCO and related compounds.  The author would like to thank Y. Koike and J.M. Tranquada for invaluable
discussions of their results.


\clearpage

\begin{table*} 
\widetext
\caption{Fundamental properties of elastic magnetic peaks of
La$_{2-x}$Sr$_{x}$Cu$_{1-y}$Zn$_{y}$O$_{4}$, La$_{1.6}$Nd$_{0.4}$Sr$_{x}$CuO$_{4}$ and La$_{2}$CuO$_{4+y}$.
 $T_{c}$ is the onset temperature of superconductivity, $T_{m}$ is the temperature the
elastic magnetic peaks appear, $\xi$ is the correlation length in the CuO$_{2}$ plane
estimated from fits to the peak profile with the resolution, and $\mu$ is the magnetic
moments of the elastic component.  The peak position is described in a polar
coordinate, namely, the incommensurability $\delta$, defined as the distance
between the magnetic peaks and the $(\frac{1}{2}\ \frac{1}{2}\ 0)$ position in the high temperature
tetragonal lattice unit, and $\theta_{Y}$, defined as the rotation of the
incommensurate magnetic peaks ($Y$ shift) with respect to the tetragonal
$[1\ 0\ 0]/[0\ 1\ 0]$ direction.}
\vspace*{0.05in}
\begin{tabular}{lrrrlldcrc}
Sample & $T_{c}$ (K) & $T_{m}$ (K) & $\xi$ (\AA) & $\delta$ (r.l.u.) & $\mu$ ($\mu_{B}$) &
$\theta_{Y}$ ($^{\circ})$ & Ref. \\
\tableline
La$_{2-x}$Sr$_{x}$Cu$_{1-y}$Zn$_{y}$O$_{4}$ \\
$\ x=0.0$             & 0    & 325 & $>600$    & 0     & 0.55 &          & \cite{Vaknin_87}\\
$\ x=0.02$            & 0    &  40 & 160$^{a}$, 25$^{b}$, 5$^{c}$ & --    & 0.3  & 45       & \cite{Matsuda_00} \\
$\ x=0.024$           & 0    &  25 &  95$^{a}$, 40$^{b}$, 3$^{c}$ & 0.016 & --   & 45       & \cite{Matsuda_00_2}\\
$\ x=0.03$            & 0    &  19 & 33        & 0.026 & 0.18 & 45       & \cite{Wakimoto_00_2,Wakimoto_unp}\\
$\ x=0.04$            & 0    &  17 & 37        & 0.037 & 0.15 & 45       & \cite{Wakimoto_00,Wakimoto_00_2,Fujita_00} \\
$\ x=0.05$            & 0    &  15 & 33$^{a}$, 25$^{b}$  & 0.045 & 0.13 & 45       & \cite{Wakimoto_00,Wakimoto_00_2} \\
$\ x=0.053$           & $<2$ &  14 & 32        & 0.045 & --   & 45       & \cite{Fujita_00} \\
$\ x=0.056$           & 6    &  13 & 39$^{D}$  & 0.051 & --   & 45       & \cite{Fujita_00} \\
                      &      &     & 31$^{P}$  & 0.049 &      & $\sim 0$ & \\
$\ x=0.06$            & 12   &  11 & 26$^{D}$  & 0.053 & 0.06 & 45       & \cite{Fujita_00,Wakimoto_00_2} \\
                      &      &     & 20$^{P}$  & 0.049 &      & $\sim 0$ & \\
$\ x=0.07$            & 17   &  11 & 27        & 0.069 & 0.05 & $\sim 0$ & \cite{Fujita_00,Wakimoto_00_2} \\
$\ x=0.10$            & 28   &  19 & 70        & 0.11  & --   & --       & \cite{Matsushita_99} \\
$\ x=0.12$            & 32   &  32 & $>200$    & 0.118 & 0.15 & 3.0      & \cite{Kimura_99,Kimura_00} \\
$\ x=0.12$, $y=0.03$  & $<5$ &  17 & 77        & 0.117 & --   & 2.6      & \cite{Kimura_99} \\
$\ x=0.13$            & 36   &  19 & 88        & 0.12  & --   & --       & \cite{Matsushita_99} \\
$\ x=0.14$, $y=0.012$ & 19   &  20 & 80        & 0.132 & --   & --       & \cite{Hirota_98} \\
$\ x=0.15$            & 37   &   0 &           &       &      &          & \cite{Yamada_98} \\
$\ x=0.18$            & 35   &   0 &           &       &      &          & \cite{Yamada_98} \\
$\ x=0.21$, $y=0.01$  & $<2$ &  20 & 70        & 0.135 & --   & 2.1      & \cite{Kimura_00} \\
\tableline
La$_{1.6-x}$Nd$_{0.4}$Sr$_{x}$CuO$_{4}$ \\
$\ x=0.12$            & 4    &  50 & $>170$    & 0.118 & 0.10 & 0$^{*}$  & \cite{Tranquada} \\
$\ x=0.15$            & 11   &  45 & --        & 0.130 & --   & --       & \cite{Tranquada} \\
$\ x=0.20$            & 15   &  20 & --        & 0.143 & --   & --       & \cite{Tranquada} \\
\tableline
La$_{2}$CuO$_{4+y}$ \\
$\ y=0.12$            & 40.5 & 40.5& $>400$    & 0.121 & 0.15 & 3.3      & \cite{Lee_99} \\
\end{tabular}
\vspace*{0.05in}
(a),(b) and (c) denote correlation lengths along the orthorhombic $a$, $b$ and $c$ axes.

(D) and (P) denote ``Diagonal'' and ``Parallel'' modulations.

(*) No shift in the $k$ direction was found for the charge-ordering peak at $(2.24\ 0\ 0)$.
\label{Table:Summary}
\narrowtext
\end{table*}


\begin{references}
\bibitem{Kastner_98} M.A. Kastner, R.J. Birgeneau, G. Shirane, and Y. Endoh, Rev.\ Mod.\ Phys.\ {\bf 70} (1998) 897  {\it
and references therein}.

\bibitem{Vaknin_87} D. Vaknin, S.K. Sinha, D.E. Moncton, D.C. Johnston, J.M. Newsom, C.R. Safenya and J.H.E. King, Phys.\
Rev.\ Lett.\ {\bf 58} (1987) 2802.

\bibitem{Sternlieb_90} B.J. Sternlieb, G.M. Luke, Y.J. Uemura, T.M. Riseman, J.H. Brewer, P.M. Gehring, K. Yamada, Y.
Hidaka, T. Murakami, T.R. Thurston and R.J. Birgeneau, Phys.\ Rev.\ B {\bf 41} (1990) 8866.

\bibitem{Keimer_92} B. Keimer, N. Belk, R.J. Birgeneau, A. Cassanho, C.Y. Chen, M.Greven, M.A. Kastner, A. Aharony, Y.
Endoh, R.W. Erwin and G. Shirane, Phys.\ Rev.\ B {\bf 46} (1992) 14034.

\bibitem{Takagi_92} H. Takagi, R.J. Cava, M. Marezio, B. Batlogg, J.J. Krajewski, W.F. Peck Jr., P. Bordet and D.E. Cox,
Phys.\ Rev.\ Lett.\ {\bf 68} (1992) 3777.

\bibitem{Nagao_93} T. Nagao, Y. Tomioka, Y. Nakayama, K. Kishino and K. Kitazawa, Phys.\ Rev.\ B {\bf 48} (1993) 9689.

\bibitem{Cheong_91} S.-W. Cheong, G. Aeppli, T.E. Mason, H. Mook, S.M. Hayden, P.C. Canfield, Z. Fisk, K.N. Clausen and
J.L. Maritinez, Phys.\ Rev.\ Lett.\ {\bf 67} (1991) 1791.

\bibitem{Yamada_98} K. Yamada, C.H. Lee, K. Kurahashi, J. Wada, S. Wakimoto, S. Ueki, H. Kimura, Y. Endoh, S. Hosoya, G.
Shirane, R.J. Birgeneau, M. Greven, M.A. Kastner and Y.J. Kim, Phys.\ Rev.\ B {\bf 57} (1998) 6165.

\bibitem{Bulut_90} N. Bulut, D. Hone, D.J. Scalapino and N.E. Bickers, Phys.\ Rev.\ Lett.\ {\bf 64} (1990) 2723.

\bibitem{Tanamoto_94} T. Tanamoto, H. Kohno and H. Fukuyama, J.\ Phys.\ Soc.\ Japan {\bf 63} (1994) 2739.

\bibitem{Tranquada} J.M. Tranquada, B.J. Sternlieb, J.D. Axe, Y. Nakamura and S. Uchida, Nature {\bf 375}, 561 (1995);
J.M. Tranquada, J.D. Axe, N. Ichikawa, Y. Nakamura, S. Uchida and B. Nachumi, Phys.\ Rev.\ B {\bf 54} (1996) 7489; J.M.
Tranquada, J.D. Axe, N. Ichikawa, A.R. Moodenbaugh, Y. Nakamura and S. Uchida, Phys.\ Rev.\ Lett.\ {\bf 78} (1997) 338;
J.M. Tranquada, N. Ichikawa and S. Uchida, Phys.\ Rev.\ B {\bf 59} (1999) 14712.

\bibitem{Emery-Kivelson} V.J. Emery, S.A. Kivelson and O. Zachar, Phys.\ Rev.\ B {\bf 56} (1997) 6120; S.A. Kivelson, E.
Frandkin and V.J. Emery, Nature {\bf 393} (1998) 550.

\bibitem{Moodenbaugh_88} A. R. Moodenbaugh {\it et al.}, Phys.\ Rev.\ B {\bf 38} (1988) 4596.

\bibitem{Axe_94} J.D. Axe and M.K. Crawford, J. Low Temp.\ Phys.\ {\bf 95} (1994) 271.

\bibitem{Goto_94} T. Goto, S. Kazama, K. Miyagawa and T. Fukase, J. Phys.\ Soc.\ Japan {\bf 63} (1994) 3494.

\bibitem{Kumagai_94} K. Kumagai, K. Kawano, I. Watanabe, K. Nishiyama and K. Nagamine, J. Supercond.\ {\bf 7} (1994) 63.

\bibitem{Hirota_98} K. Hirota, K. Yamada, I. Tanaka and H. Kojima, Physica B, {\bf 241-243} (1998) 817.

\bibitem{Suzuki_98} T. Suzuki, T. Goto, K. Chiba, T. Shinoda, T. Fukase, H. Kimura, K. Yamada, M. Ohashi and Y. Yamaguchi,
Phys.\ Rev.\ B {\bf 57} (1998) 3229.

\bibitem{Kimura_99} H. Kimura, K. Hirota, H. Matsushita, K. Yamada, Y. Endoh, S.H. Lee, C.F. Majkrzak, R.W. Erwin, G.
Shirane, M. Greven, Y.S. Lee, M.A. Kastner and R.J. Birgeneau, Phys.\ Rev.\ B {\bf 59} (1999) 6517.

\bibitem{Kimura_00} H. Kimura, H. Matsushita, K. Hirota, Y. Endoh, K. Yamada, G. Shirane, Y.S. Lee, M.A. Kastner and R.J.
Birgeneau, Phys.\ Rev.\ B {\bf 61} (2000) 14366.

\bibitem{Matsushita_99} H. Matsushita, H. Kimura, M. Fujita, K. Yamada, K. Hirota and Y. Endoh, J. Phys.\ Chem.\ Solids
{\bf 60} (1999) 1071.

\bibitem{Lee_99} Y.S. Lee, R.J. Birgeneau, M.A. Kastner, Y. Endoh, S. Wakimoto, K. Yamada, R.W. Erwin, S.H. Lee and G.
Shirane, Phys.\ Rev.\ B {\bf 60} (1999) 3643.

\bibitem{Wakimoto_99} S. Wakimoto, G. Shirane, Y. Endoh, K. Hirota, S. Ueki, K. Yamada, R.J. Birgeneau, M.A. Kastner, Y.S.
Lee, P.M. Gehring and S.H. Lee, Phys.\ Rev.\ B {\bf 60} (1999) R769.

\bibitem{Wakimoto_00} S. Wakimoto, R.J. Birgeneau, M.A. Kastner, Y.S. Lee, R.W. Erwin, P.M. Gehring, S.H. Lee, M. Fujita,
K. Yamada, Y. Endoh, K. Hirota and G. Shirane, Phys.\ Rev.\ B {\bf 61} (2000) 3699.

\bibitem{Wakimoto_00_2} S. Wakimoto, R.J. Birgeneau, Y.S. Lee and G. Shirane (unpublished).

\bibitem{Wakimoto_unp} S. Wakimoto (private communication).

\bibitem{Fujita_00} M. Fujita, K. Yamada, H. Hiraka, S.H. Lee, P.M. Gehring, S. Wakimoto and G. Shirane (unpublished).

\bibitem{Matsuda_00} M. Matsuda, R.J. Birgeneau, P. B\"{o}ni, Y. Endoh, M. Greven, M.A. Kastner, S.H. Lee, Y.S. Lee, G.
Shirane, S. Wakimoto and K. Yamada, Phys.\ Rev.\ B {\bf 61} (2000) 4326.

\bibitem{Matsuda_00_2} M. Matsuda, M. Fujita, K. Yamada, R.J. Birgeneau, M.A. Kastner, H. Hiraka, Y. Endoh, S. Wakimoto
and G. Shirane, Phys.\ Rev.\ B (2000) Ocober 1.

\bibitem{Kimura_00_2} H. Kimura, K. Hirota, M. Aoyama, T. Adachi, T. Kawamata, Y. Koike, K. Yamada and Y. Endoh, {\it
Proceedings for ASR2000}  

\bibitem{Nachumi_96} B. Nachumi, A. Keren, K. Kojima, M. Larkin, G.M. Luke, J. Merrin, O. Tchernysh\"{o}v, Y.J. Uemura, N.
Ichikawa, M. Goto and S. Uchida, Phys.\ Rev.\ Lett.\ {\bf 77} (1996) 5421.
\end{references}
\end{document}